\documentclass[twocolumn]{webofc}
\usepackage{tikz}
\usepackage{graphicx}
\usepackage{epsfig}

\newcommand{\preliminary}[2][3]{
	\begin{tikzpicture}
	\node[anchor=south west,inner sep=0] (murks) at (0,0) {#2};
	\begin{scope}[x={(murks.south east)},y={(murks.north west)}]
	\node [rotate=30,scale=#1,text opacity=0.2]
	at (murks.center) {Preliminary};
	\end{scope}
	\end{tikzpicture}
}
\usepackage[varg]{txfonts}   
\begin{document}
\title{Strangeness Photoproduction at the BGO-OD experiment}
%
%

\author{\firstname{Thomas} \lastname{Jude}\inst{1}\fnsep\thanks{\email{jude@physik.uni-bonn.de}}\and
\firstname{Stefan} \lastname{Alef}\inst{1}\and
\firstname{Patrick} \lastname{Bauer}\inst{1}\and
\firstname{Reinhard} \lastname{Beck}\inst{2}\and
\firstname{Alessandro} \lastname{Braghieri}\inst{13}\and
\firstname{Philip} \lastname{Cole}\inst{3}\and
\firstname{Rachele} \lastname{Di Salvo}\inst{4}\and
\firstname{Daniel} \lastname{Elsner}\inst{1}\and
\firstname{Alessia} \lastname{Fantini}\inst{4,5}\and
\firstname{Oliver} \lastname{Freyermuth}\inst{1}\and
\firstname{Francesco} \lastname{Ghio}\inst{6,7}\and
\firstname{Anatoly} \lastname{Gridnev}\inst{8}\and
\firstname{Daniel} \lastname{Hammann}\inst{1}\and \newline
\firstname{J\"urgen} \lastname{Hannappel}\inst{1}\and 
\firstname{Katrin} \lastname{Kohl}\inst{1}\and
\firstname{Nikolay} \lastname{Kozlenko}\inst{8}\and
\firstname{Alexander} \lastname{Lapik}\inst{9}\and
\firstname{Paolo} \lastname{Levi Sandri}\inst{10}\and
\firstname{Valery} \lastname{Lisin}\inst{9}\and \newline
\firstname{Giuseppe} \lastname{Mandaglio}\inst{11,12}\and
\firstname{Roberto} \lastname{Messi}\inst{4,5}\and
\firstname{Dario} \lastname{Moricciani}\inst{4}\and
\firstname{Vladimir} \lastname{Nedorezov}\inst{9}\and
\firstname{Dmitry} \lastname{Novinsky}\inst{8}\and \newline
\firstname{Paolo} \lastname{Pedroni}\inst{13}\and
\firstname{Andrei} \lastname{Polonski}\inst{9}\and
\firstname{Bj\"orn-Eric} \lastname{Reitz}\inst{1}\and
\firstname{Mariia} \lastname{Romaniuk}\inst{4}\and
\firstname{Georg} \lastname{Scheluchin}\inst{1}\and
\firstname{Hartmut} \lastname{Schmieden}\inst{1}\and
\firstname{Victorin} \lastname{Sumachev}\inst{8}\and
\firstname{Viacheslav} \lastname{Tarakanov}\inst{8}\and
\firstname{Christian} \lastname{Tillmanns}\inst{1}}

\institute{Rheinische Friedrich-Willhelms-Universit\"at Bonn, Physikalisches Institut, Nu\ss allee 12, 53115 Bonn, Germany \and 
Helmholtz-Institut fuer Strahlen- und Kernphysik, Universitaet Bonn, Nussallee 1-16, D-53115 Bonn Germany \and
Lamar University, Department of Physics, Beaumont, Texas, 77710, USA \and
INFN Roma ``Tor Vergata", Rome, Italy \and
Università di Roma ``Tor Vergata", Via della Ricerca Scientifica 1, 00133 Rome, Italy \and
INFN sezione di Roma La Sapienza, P.le Aldo Moro 2, 00185 Rome, Italy \and
Istituto Superiore di Sanita, Viale Regina Elena 299, 00161 Rome, Italy \and
Petersburg Nuclear Physics Institute, Gatchina, Leningrad District, 188300, Russia \and
Russian Academy of Sciences Institute for Nuclear Research, prospekt 60-letiya Oktyabrya 7a, Moscow 117312, Russia \and
INFN - Laboratori Nazionali di Frascati, Via E. Fermi 40, 00044 Frascati, Italy \and
INFN sezione Catania, 95129 Catania, Italy \and
Universita degli Studi di Messina, Via Consolato del Mare 41, 98121 Messina, Italy\and
\,INFN sezione di Pavia, Via Agostino Bassi, 6 - 27100 Pavia, Italy}

\abstract{

	
	The BGO-OD experiment at the ELSA accelerator facility uses an energy tagged bremsstrahlung photon beam to investigate the excitation structure of the nucleon. The setup consists of a highly segmented BGO calorimeter surrounding the target, with a particle tracking magnetic spectrometer at forward angles.
	
	BGO-OD is ideal for investigating low momentum transfer processes due to the acceptance and high momentum resolution at forward angles. In particular, this enables the investigation of strangeness photoproduction where t-channel exchange mechanisms play an important role.  This also allows access to low momentum exchange kinematics where extended, molecular structure may manifest in reaction mechanisms.  
	
	First key results at low $t$ indicate a cusp-like structure in $K^+\Sigma^0$ photoproduction at $W = 1900$\,MeV, line shapes and differential cross sections for $K^+\Lambda$(1405)$\rightarrow K^+\Sigma^0\pi^0$, and a peak structure in $K^0_S\Sigma^0$ photoproduction.  The peak in the $K^0_S\Sigma^0$ channel appears consistent with meson-baryon generated states, where equivalent models have been used to describe the $P_C$ pentaquark candidates in the heavy charmed quark sector.
}
\maketitle
\section{Introduction}
\label{intro}

Hadron spectroscopy has been used for many years to understand  the interactions of the constituents and the relevant degrees of freedom. 
Despite the wealth of data available using both pion and photon beams and a multitude of final states, there are still many  resonances predicted by constituent quark models (CQMs) that have not been observed experimentally (see for example Ref.~\cite{loering01}).  This is in spite of major advances in partial wave analysis working in tandem with experimental research (see for example Ref.~\cite{anisovich14}).  Even some lowest lying states are difficult to reconcile within a CQM.  The pattern of the mass and parity of the $N(1440)1/2^+$ and $N(1535)1/2^-$ for example are hard to describe assuming valence quarks in a mutually generated potential.  In the strange sector, the mass ordering of the $\Lambda(1405)$ compared to the $N^*(1535)$ (despite the $\Lambda(1405)$ being a $uds$ singlet state) and the mass difference to the $\Lambda(1520)$ spin orbit partner is hard to understand within a CQM framework.

Since the conception of the quark model, there have been descriptions of hadrons with more than three constituent quarks~\cite{gellmann64, jaffe77, strottmann79}, and models explicitly including light mesons interacting as elementary objects have had markedly improved success in describing nucleon excitation spectra~\cite{glozman96, recio04, lutz04}.
Such interactions may give rise to meson rescattering effects near production thresholds, and molecular like systems.

The BGO-OD experiment at the ELSA facility, Bonn, Germany, is a unique setup that is ideal to study molecular like interactions in strangeness photoproduction which may manifest in the reaction mechanisms.  This proceedings is as follows:  section~\ref{sec:parallels} discusses parallels between the heavy charm and lighter strange quark sector, where there maybe equivalent evidence of molecular-like systems.  The BGO-OD experiment and performance is described in sections~\ref{sec:bgood} and \ref{sec:performance}, and the first preliminary results in associated strangeness photoproduction are shown in section~\ref{sec:strangeresults}.

\section{Exotic hadron structure - parallels in strange and charmed quark sectors?}\label{sec:parallels}

In the heavy, charmed quark sector, the pentaquark candidates, $P_C(4450)$ and $P_C(4380)$ were one of the first indications of baryons of more than three constituent quarks~\cite{aaij15}.  It is still an open question to whether these are five quark bound systems, or a molecular like system to some extent, with models successfully describing them as meson-baryon dynamically generated states (see for example, Ref.~\cite{wu10b}).
In the meson sector, the $X(3872)$ close to the $D^0\bar{D}^{0*}$ threshold has also been described as a molecular state~\cite{tornqvist04}.

In the lighter strange sector, models including meson-baryon interactions have had improved success.  The $\Lambda(1405)$, lying close to the $\bar{K}N$ threshold appears to be dynamically generated to some extent~\cite{nacher03, molina16,jido03}, which is supported by Lattice QCD calculations~\cite{hall15}.  A cusp-like structure observed in the $K^0_S\Sigma^+$ differential cross section at the $K^*$ thresholds~\cite{ewald12} has been successfully described by the destructive interference between two vector meson-baryon dynamically generated resonances~\cite{gonzalez09, sarkar10, oset10}.
There are remarkable comparisons to the heavier charmed pentaquarks described by an equivalent model~\cite{wu10b}.   Both share quantum numbers, $J^P = 3/2^-$, equivalent pion exchange transitions and close to two and three body thresholds.

\begin{figure}[hbt]
	\centering
	{\includegraphics[width=\columnwidth,trim={0cm 0 0 0},clip=true]{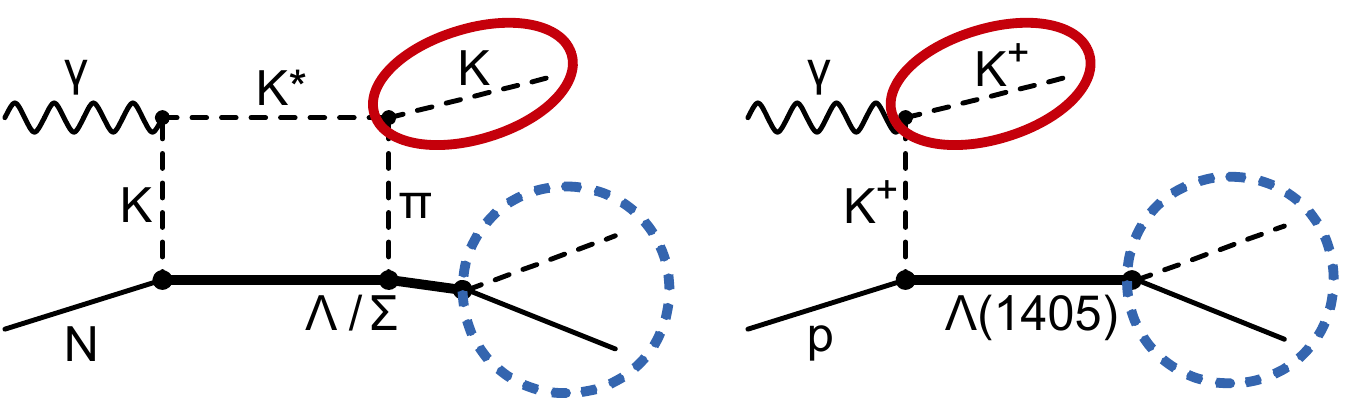}}
	\caption{Example of kinematics ideally suited for BGO-OD.}
	\label{fig:forwardkinematics}       
\end{figure}

To study systems where molecular like interactions may manifest, it is vital to access kinematic regions where the residual baryon remains at low momentum.  Using a real, energy tagged photon beam, this corresponds to very forward meson angles.
BGO-OD is ideal for studying such systems in associated strangeness photoproduction.
The experiment comprises a forward spectrometer for charged particle identification at forward angles, coupled with a central calorimeter ideal to identify hyperon decays.
Fig.~\ref{fig:forwardkinematics} depicts two examples where the kinematics suit this setup.  In in both scenarios, the $K^+$ travels forwards, leaving either a residual $K^*Y$ system (left) or $\Lambda$(1405) (right) at low momentum to decay almost isotropically.
With the ability to reconstruct complicated mixed charge final states, BGO-OD is complementary to other facilities such as CBELSA-TAPS, Crystal Ball and CLAS.

\section{The BGO-OD experiment at ELSA}\label{sec:bgood}

\begin{figure*}[hbt]
	\centering
	{\includegraphics[width=\textwidth,trim={0cm 0 0 4cm},clip=true]{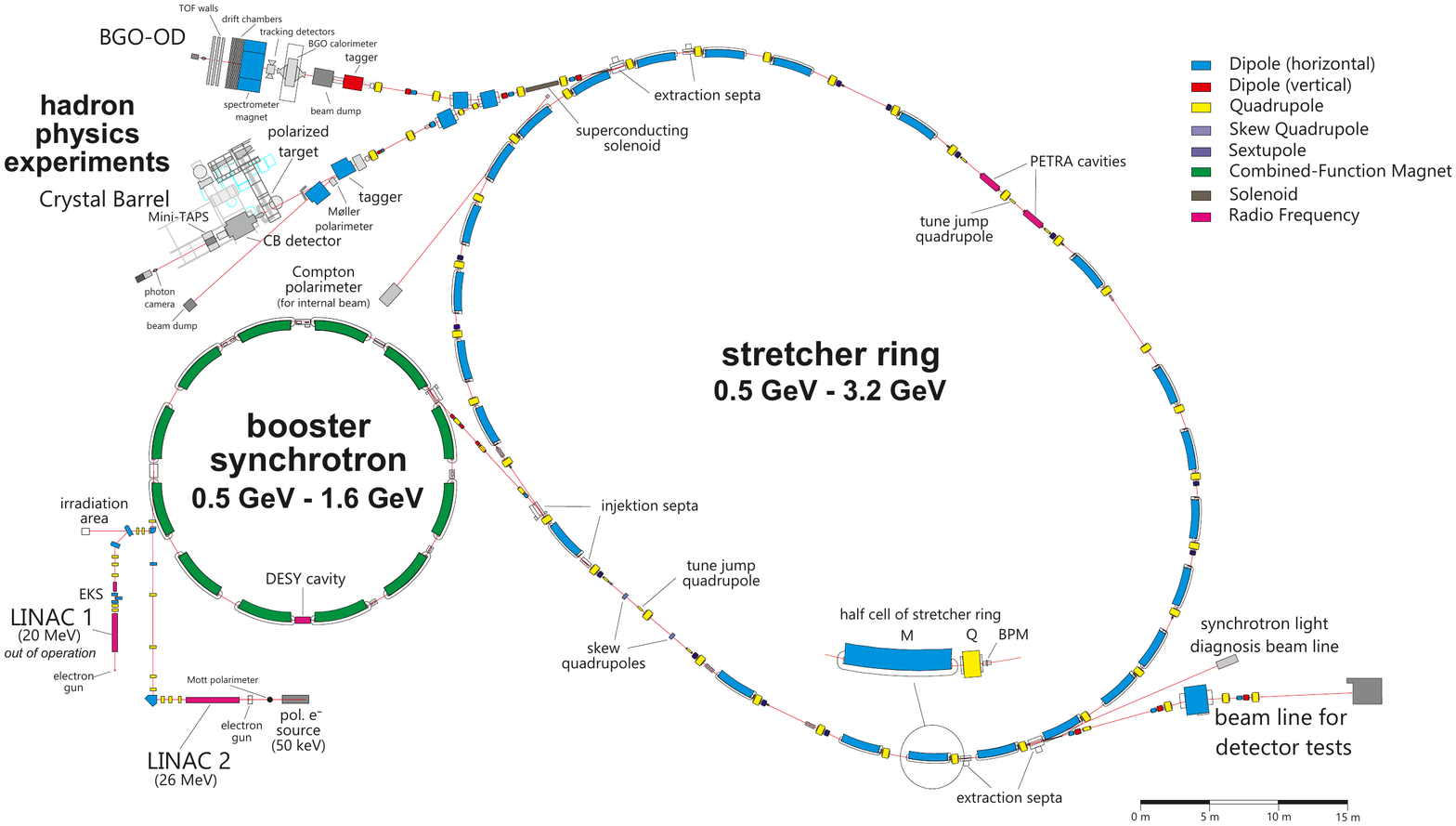}}
	\caption{Overview of the ELSA accelerator facility.  Figure adapted from Ref.~\cite{hillert17}.}
	\label{fig:elsa}       
\end{figure*}

The ELectron Stretcher Accelerator (ELSA)~\cite{hillert06}, shown in Fig.~\ref{fig:elsa} is situated at the University of Bonn's Physikalisches Institut.  ELSA is comprised of three parts.  A LINAC delivers a 20~MeV electron beam to a booster synchrotron which accelerates this pulsed beam up to 1.6~MeV.  This enters the stretcher ring, whereupon it is accelerated up to 3.2~GeV and accumulated to provide a 100\% duty factor continuous wave beam.  The beam is subsequently delivered to one of three experimental areas:  a beam line for detector tests (bottom right), the CBELSA/TAPS experiment~\cite{Aker:1992ny,Gabler:1994ay,Novotny:1991ht} or the BGO-OD experiment (both top left).

\begin{figure*}[hbt]
\centering
\includegraphics[width=\textwidth,clip]{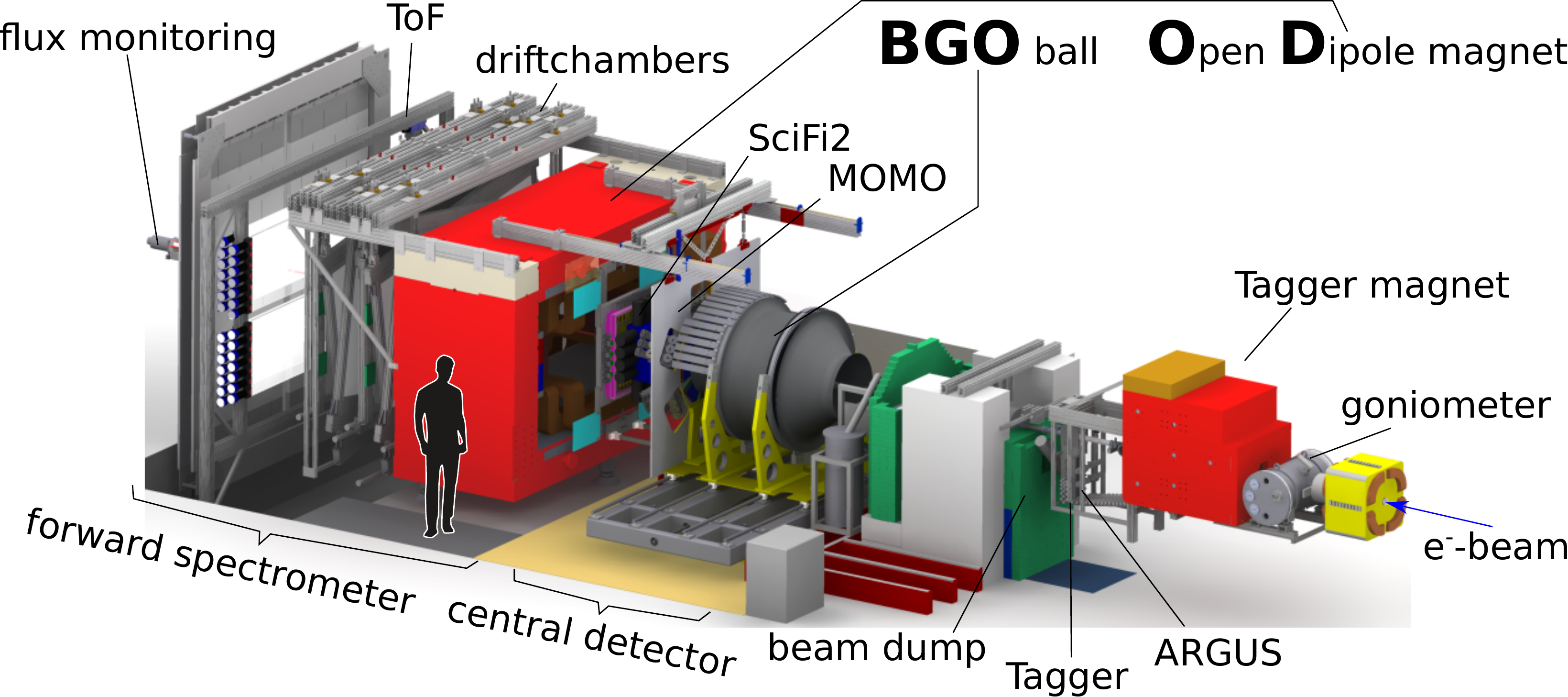}
\caption{The BGO-OD experiment}
\label{fig:bgood}       
\end{figure*}

The BGO-OD experiment, shown in Fig.~\ref{fig:bgood}, is comprised of two main components.  A central region, consisting of a calorimeter and charged particle tracking detectors, and a forward spectrometer for charged particle identification and momentum reconstruction.

The electron beam from ELSA impinges upon a thin radiator (bottom right of Fig.~\ref{fig:bgood} within the goniometer tank) producing bremsstrahlung photons.  A metal or crystal radiator can be used to produce either unpolarised (incoherent) or polarised (coherent) bremsstrahlung photons respectively.  Photon energies are determined by momentum analysing the post-bremsstrahlung electrons in the Photon Tagger, over a range of 10\% to 90\% of incoming electron beam energy.

\begin{figure}[h]
	\centering
	\includegraphics[width=\columnwidth,clip]{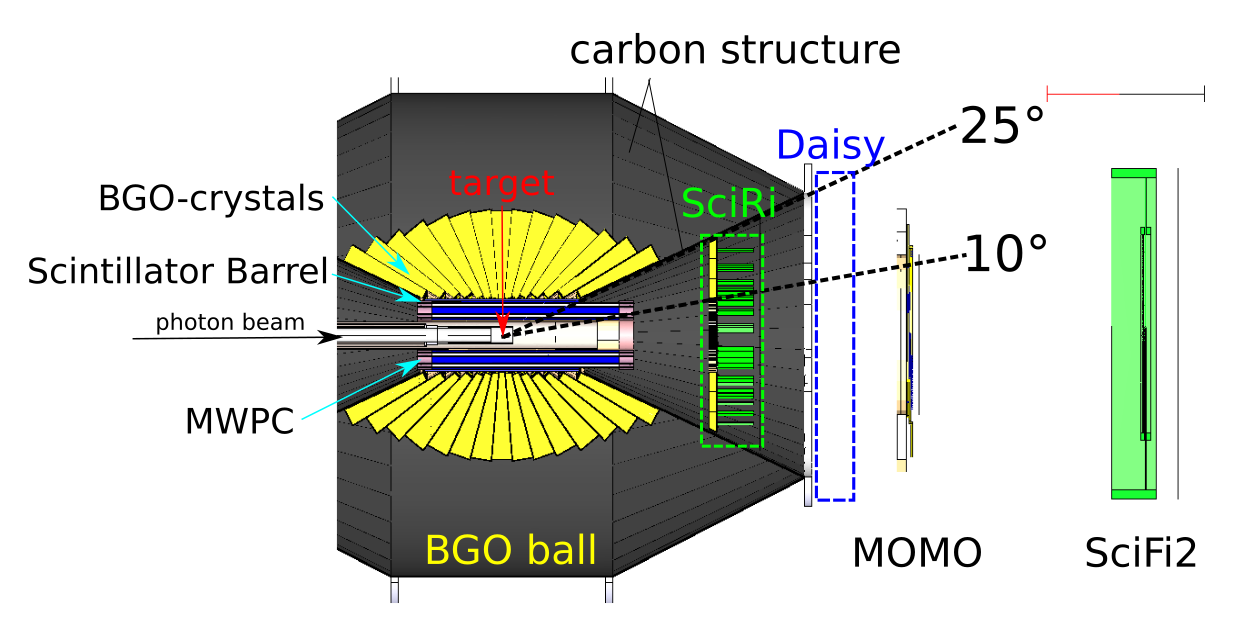}
	\caption{Slice view of the central detector consisting of the segmented BGO Rugby Ball, cylindrical inner scintillator barrel and MWPC, and
		the SciRi scintillator ring detector in forward direction. Also depicted are MOMO and SciFi, the front tracking fibre detectors
		of the forward spectrometer and the MRPC detector under construction.}
	\label{fig:bgoodsliceview}       
\end{figure}

The photon beam is subsequently collimated and incident upon the target at the centre of the BGO Rugby Ball.  Liquid hydrogen, liquid deuterium and carbon targets have previously been used.  Fig.~\ref{fig:bgoodsliceview} shows a sliced view of the central calorimeter region.  A barrel of 32 plastic scintillators surrounds the target to veto between charged and neutral particles and to provide charged particle identification via $\Delta E-E$ techniques.  Two MultiWire Proportional Chambers (MWPCs) surround this for accurate determination of charged particle trajectories, reaction and decay vertices.
The BGO Rugby Ball, consisting of 480 BGO crystals, covers the polar angle range of $25^\circ - 155^\circ$.  Each crystal is 24~cm long, corresponding to 21 radiation lengths.  Separate time readout per crystal and a resolution of approximately $2-3$~ns enable clean identification of neutral meson decays.

The small intermediate range between the BGO Rugby Ball and the Forward Spectrometer is covered by the Scintillating Ring detector (SciRi).  SciRi is comprised of three rings of 32 scintillating plastic elements each to measure directions of charged particles.

The Forward Spectrometer is a combination of tracking detectors, open dipole magnet and Time of Flight Walls (ToF Walls).  The Scintillating fibre detectors, MOMO and SciFi, track particles from the target before their trajectories are curved in the magnetic field of the open dipole magnet, with an integrated field of $\int B\mathrm{d}l \approx 0.71$~Tm.  The acceptance of the forward spectrometer covers polar angles of 12$^\circ$ and 8$^\circ$ in the horizontal and vertical directions respectively.

Eight double-sided drift chambers trace particle trajectories downstream from the magnetic field.  Each drift chamber is orientated to measure either horizontal, vertical or $\pm 9^\circ$ from horizontal information (named X,Y,U and V respectively).
Using accurate three dimensional maps of the magnetic field (both measured and simulated), particle trajectories and momenta are determined using combinations of all tracking detectors, with a resolution of $\delta p/p \approx 3\%$.
Three ToF Walls measure the time of incident particles relative to the reaction start time.
Particle $\beta$ is determined after accounting for the trajectory length and energy loss en route.  

Two photon monitoring detectors, GIM and Flumo are located behind the TOF walls to measure absolute photon flux.

For a detailed description of the BGO-OD setup and running conditions see Ref.~\cite{technicalpaper}.

\section{BGO-OD performance}
\label{sec:performance}

\begin{figure}[h]
	\centering
	\preliminary{\includegraphics[width=\columnwidth,trim={10cm 0 0 0},clip=true]{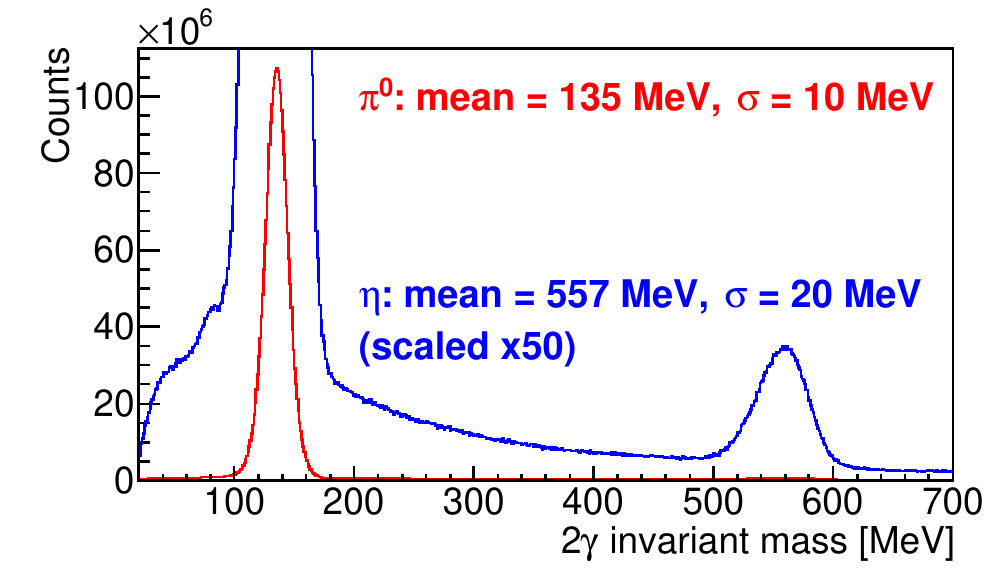}}
	\caption{Neutral meson identification using the BGO Rugby Ball.  The two photon invariant mass spectrum has peaks corresponding to the accepted $\pi^0$ and $\eta$ masses.  Approximate mean and sigma values given inset.}
	\label{fig:neutralmeson}       
\end{figure}

Fig.~\ref{fig:neutralmeson} shows the invariant mass of two photons identified in the BGO Rugby Ball after selecting a recoiling missing mass consistent with that of a proton.   A clean identification of $\pi^0$ and $\eta$ mesons can be made

\begin{figure}[h]
	\centering
	 \preliminary{\includegraphics[width=\columnwidth,trim={0cm 0 0 0},clip=true]{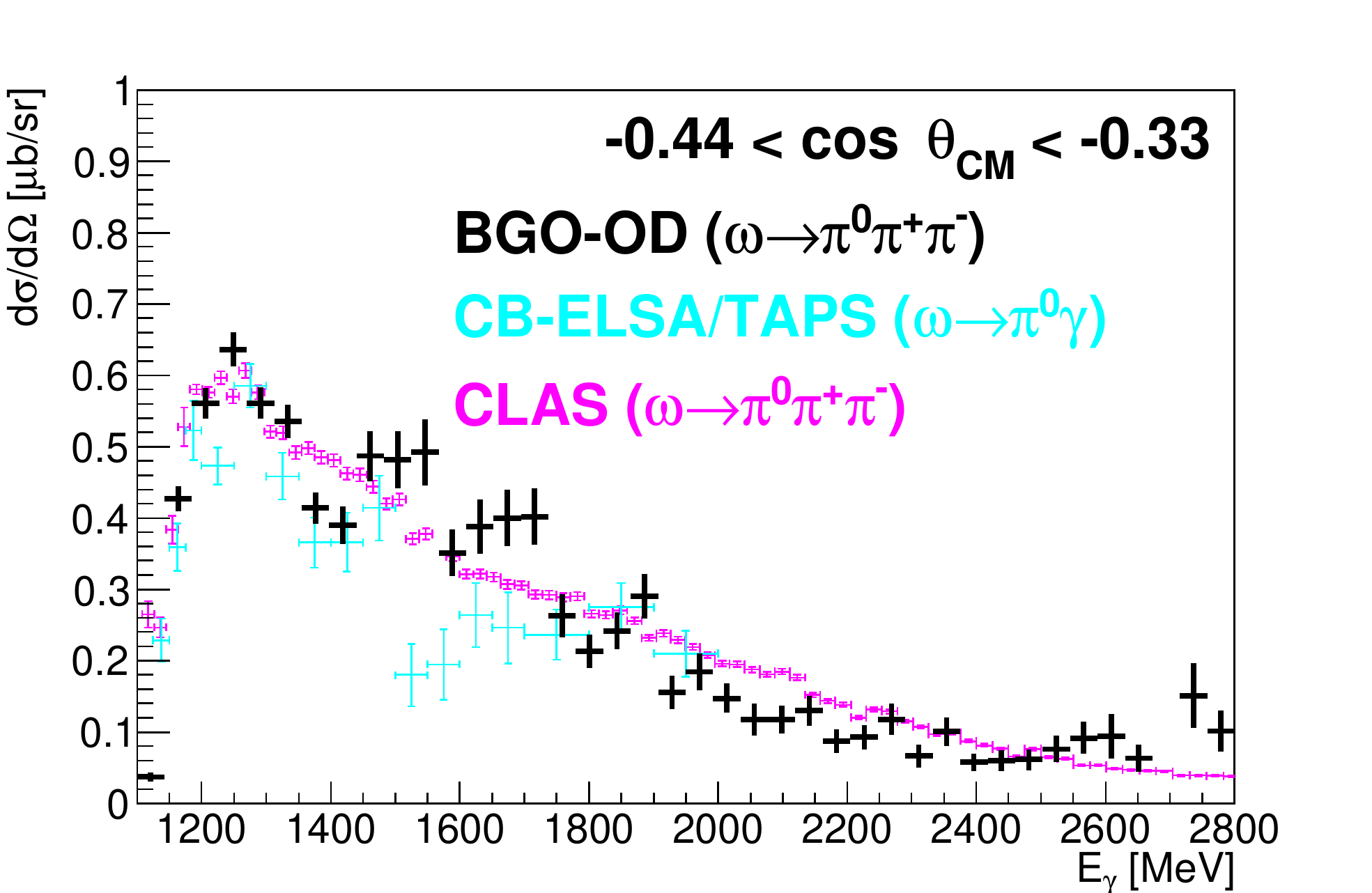}}\\
		 \preliminary{\includegraphics[width=\columnwidth,trim={0cm 0 0 0},clip=true]{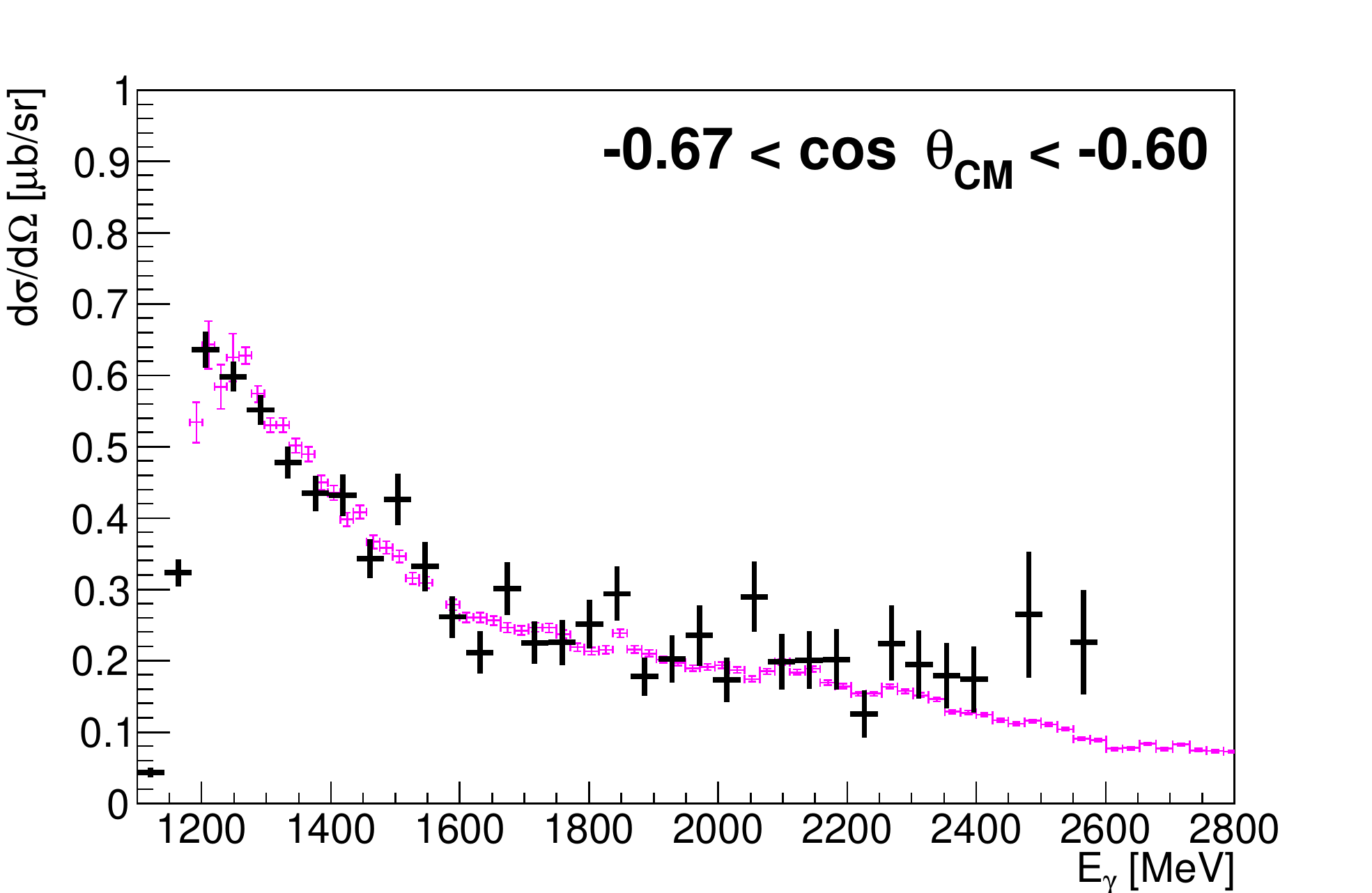}}
	\caption{Preliminary $\gamma p \rightarrow \omega p$ differential cross sections identified via the mixed charge decay, $\omega \rightarrow \pi^+\pi^-\pi^0$ (black points).  For comparison, data from Dietz \textit{et al.}~\cite{dietz15} and Williams \textit{et al.}~\cite{williams09}  are shown in blue and magenta respectively.  Analysis by G. Scheluchin~\cite{georgthesis}.}
	\label{fig:omegacs}       
\end{figure}

Numerous "bench mark" cross section measurements have been made to ensure the accuracy of the simulated geometry, trigger conditions, photon beam flux normalisation and event reconstruction (including kinematic fitting algorithms).  One example is shown in Fig.~\ref{fig:omegacs}, where the differential cross section for $\gamma p \rightarrow \omega p$ was determined with a limited data set.  The data are consistent with previous measurements.  This analysis employed the unique mixed charge identification available at BGO-OD, where the decay $\omega \rightarrow \pi^0 \pi^+\pi^-$ was identified.

\begin{figure}[h]
	\centering
		\preliminary{\includegraphics[width=\columnwidth,clip=true]{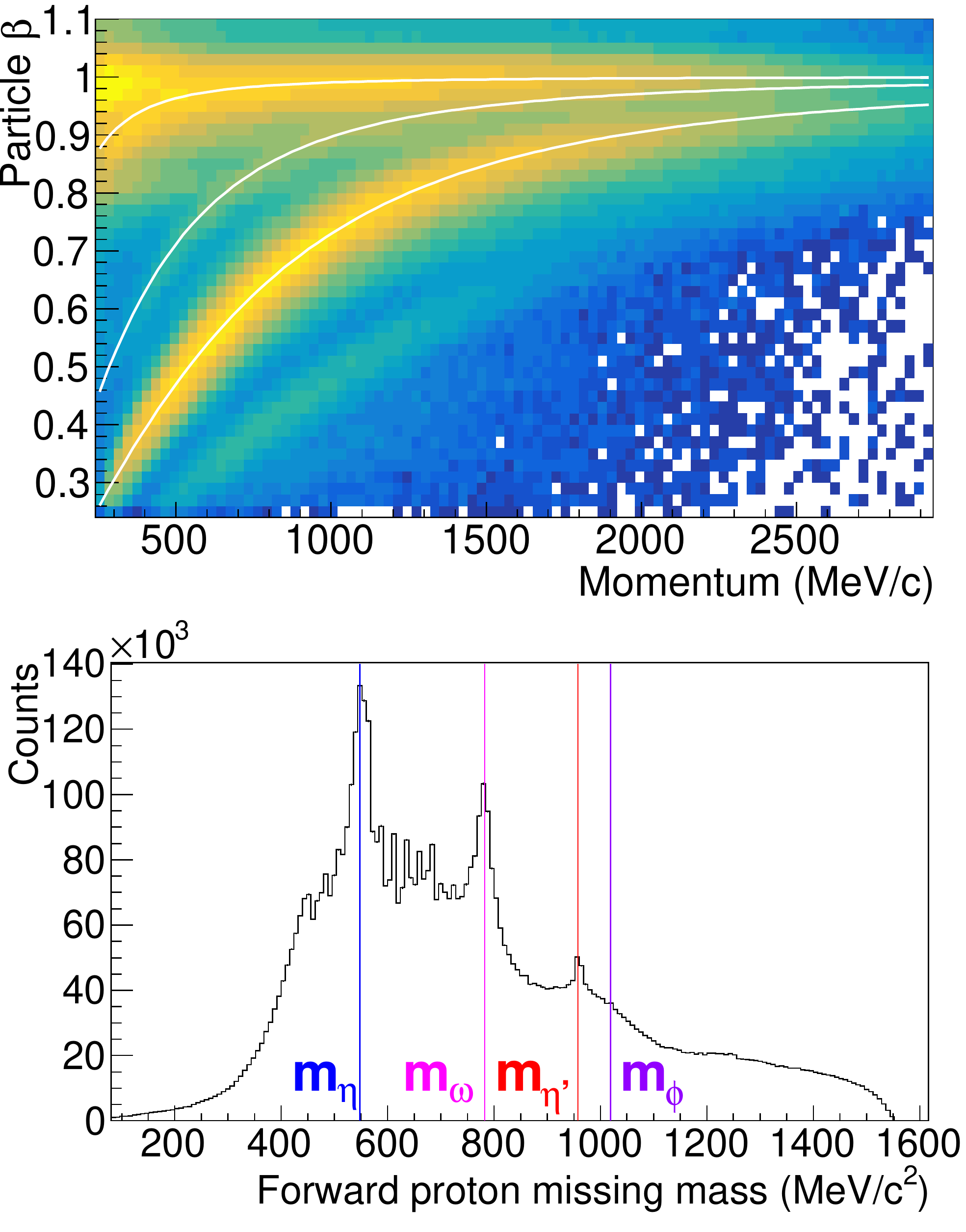}}
	\caption{Particle identification in the forward spectrometer.  Top:  $\beta$ versus momentum.  Characteristic loci for $\pi^+$, protons and $K^+$ are indicated by white lines.
Bottom:  Mass recoiling from forward particles.  Peaks from single meson photoproduction indicated.}
	\label{fig:forspec}       
\end{figure}

Fig.~\ref{fig:forspec} demonstrates particle and momentum determination using the forward spectrometer.  Particle $\beta$ versus momentum yields characteristic loci for particle species (top).  After selecting forward going protons, and with no additional selection criteria, reaction channels can be easily identified.  The bottom plot shows the missing mass recoiling from the forward proton, where peaks corresponding to single meson photoproduction are evident.
The smooth distribution under the peaks originates from multiple mesons in the final state.

\section{Preliminary results in strangeness photoproduction}\label{sec:strangeresults}

The results shown below are from the first stage of an extensive strangeness photoproduction programme.
Channels with both neutral and charged kaons are investigated, using both liquid hydrogen and deuterium targets.  The forward spectrometer for charged particle identification and calorimeter for neutral meson and hyperon decays yield unique kinematics complementary to existing facilities.

\subsection{Neutral kaon photoproduction}\label{sec:k0}

BGO-OD is able to identify $K^0_S$ via both the neutral and charged pion decays.
The model of Ramos and Oset~\cite{ramos13}, which described the cusp in $\gamma  p \rightarrow K^0_S\Sigma^+$ as the destructive interference of two dynamically generated resonances predicts the same states will constructively interfere in $\gamma n \rightarrow K^0_S\Sigma^0$, resulting in a peak in the cross section.  BGO-OD is ideally suited to search for this "smoking gun" signal.

\begin{figure}[h]
	\centering
	\preliminary{\includegraphics[width=\columnwidth,clip=true]{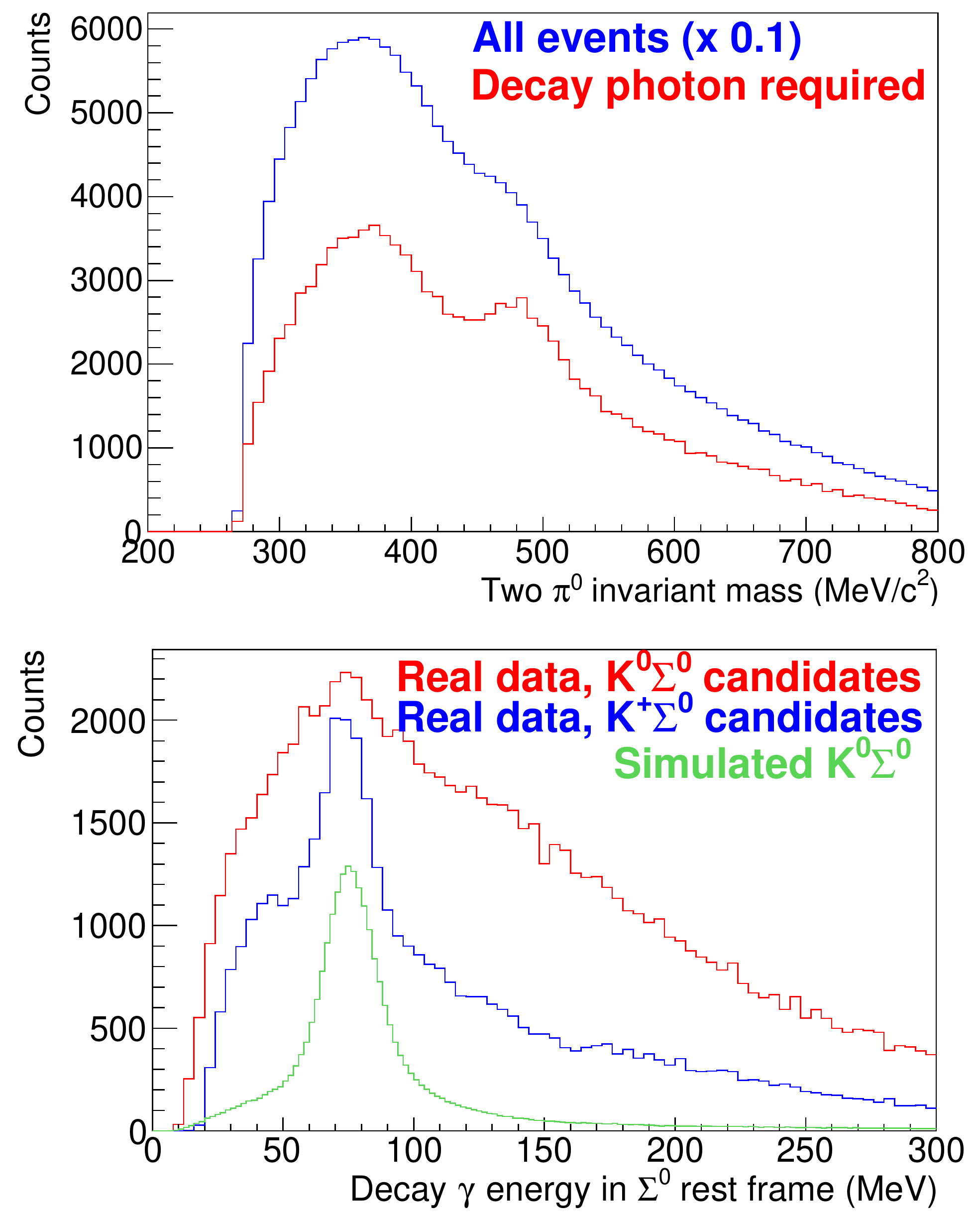}}
	\caption{Top: Identification of $\gamma n \rightarrow K^0_S\Sigma^0$ via $K^0_S\rightarrow 2\pi^0$, with and without the $\Sigma^0\rightarrow \gamma \Lambda$ decay additionally identified (red and blue lines respectively).
	Bottom: Energy of $\Sigma^0$ decay photon candidates in the $\Sigma^0$ rest frame for different scenarios labelled inset.  Analysis by K. Kohl~\cite{katrinthesis}.}
	\label{fig:sigma0decay}       
\end{figure}

The reaction $\gamma p \rightarrow K_S^0 \Sigma^0$ is identified via $K^0_S\rightarrow 2\pi^0$ in the BGO Rugby Ball and selecting a missing mass consistent with $\Sigma^0$.
A peak at the $K^0_S$ invariant mass can be seen in Fig.~\ref{fig:sigma0decay}(top), above a broad distribution of background channels.  This is dominantly uncorrelated three pion photoproduction yielding the same final state.
To improve the signal to background ratio, the photon from the decay $\Sigma^0 \rightarrow \Lambda\gamma$ was additionally identified in the BGO Rugby Ball.  After the photon momentum is boosted into the rest frame of the $\Sigma^0$, the fixed energy is close to the $\Sigma^0 - \Lambda$ mass difference of 77~MeV.  This is shown in Fig.~\ref{fig:sigma0decay} (bottom), for simulated data (where the peak is immediately evident) and real data, where a small peak resides over a large background.  To demonstrate the veracity of this technique, the channel $K^+\Sigma^0$ was used, where there is a much cleaner signal.   The peak is consistent with the other data.
After selecting events around this peak, the invariant mass of the $K^0_S$ is more prominent above background.

To remove $\gamma p \rightarrow K^0_S\Sigma^+$ events from the yield, an identical analysis was performed using liquid hydrogen target data.  The kinematics were smeared according to the momentum distribution of nucleons in deuterium,
and the signal subtracted from the $K^0_S$ invariant mass plot.

The selection of the $\Sigma^0\rightarrow \gamma \Lambda$ decay removed almost all $\gamma n \rightarrow K^0_S \Lambda$ events.  The remaining background was subtracted by scaling the known cross section~\cite{compton17} according to the measured efficiency and photon flux.  This was approximately 16\,\% and 2\,\% of the yield at photon beam energies 1250\,MeV and 1750\,MeV respectively.

\begin{figure}[h]
	\centering
	\preliminary{\includegraphics[width=\columnwidth,clip=true]{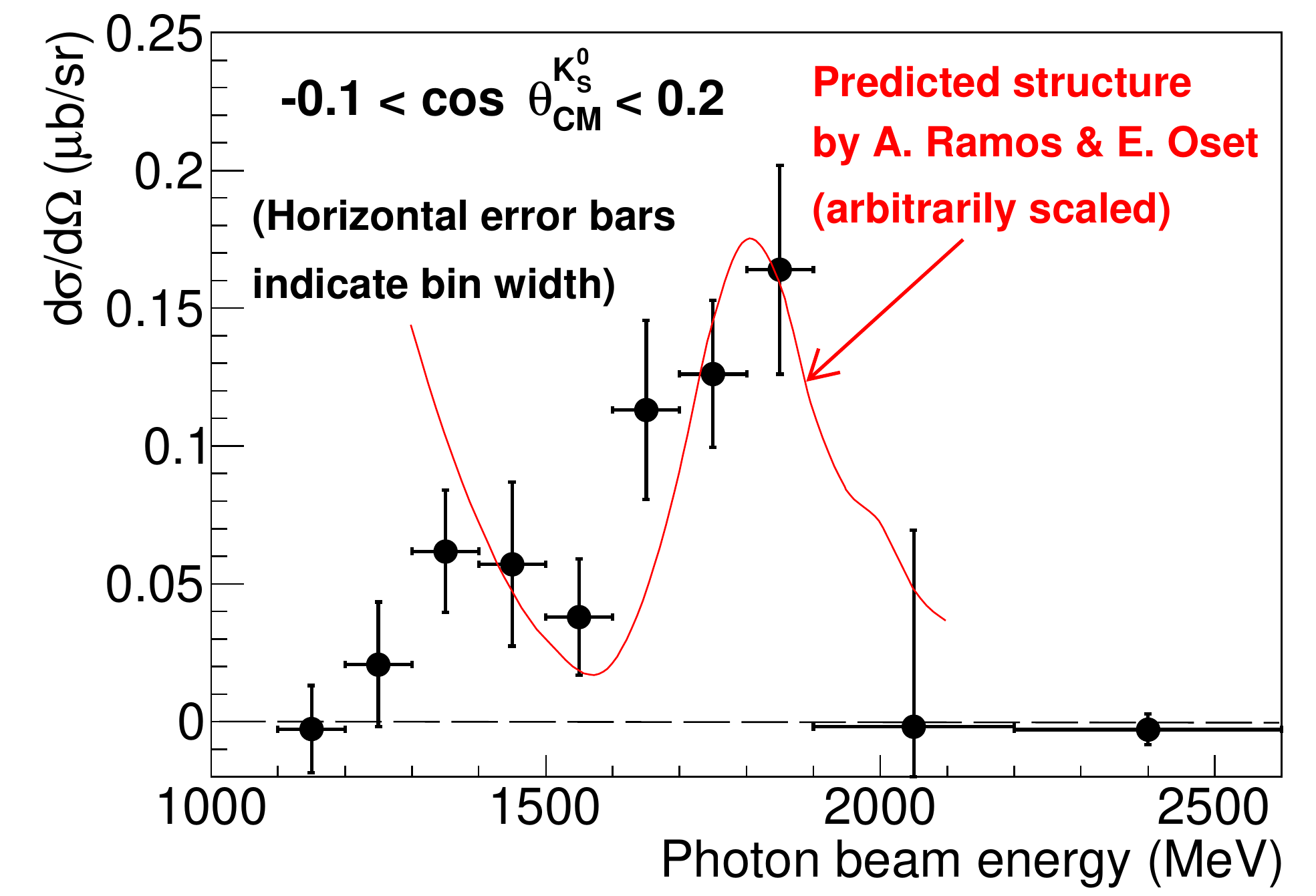}}
	\caption{Preliminary $\gamma p \rightarrow K^0_S \Sigma^0$ differential cross section.  The arbitrarily scaled  peaked structure predicted by Ramos and Oset~\cite{ramos13} is overlayed (see text for details).  Analysis by K. Kohl~\cite{katrinthesis}.}
	\label{fig:k0sigma0cs}       
\end{figure}

Fig.~\ref{fig:k0sigma0cs} shows an example of a preliminary differential cross section for $K^0_S\Sigma^0$ photoproduction.  This shows a peak structure at approximately 1700-1800~MeV photon beam energy, reducing dramatically to the point where the signal is negligible above 2~GeV.  The peak structure  predicted by Oset and Ramos~\cite{ramos13}, arising from the coherent interference of the vector-meson dynamically generated resonances appears to be consistent.  It should be noted that the model prediction is for the integrated cross section which has been scaled by a factor of five to approximately match the data.  This is only to demonstrate that the shape of the structure appears consistent, but with no constraints upon the absolute scale.  For further details see Ref.~\cite{katrinNSTAR}.

A complimentary analysis to identify the decay \newline $K^0_S\rightarrow \pi^+\pi^-$ is underway~\cite{Bjoernthesis}, where it is anticipated that more forward angles can be reached.  This analysis uses the excellent spatial resolution afforded by the MWPC to identify the $K^0_S$ decay vertex outside of the target fiducial volume ($c\tau \approx 2.7$~cm).

\subsection{$K^+ (\Lambda(1405) \rightarrow \pi^0\Sigma^0)$ photoproduction - Lineshapes and differential cross sections}

\begin{figure}[h]
	\centering
	\preliminary{\includegraphics[width=\columnwidth,clip=true]{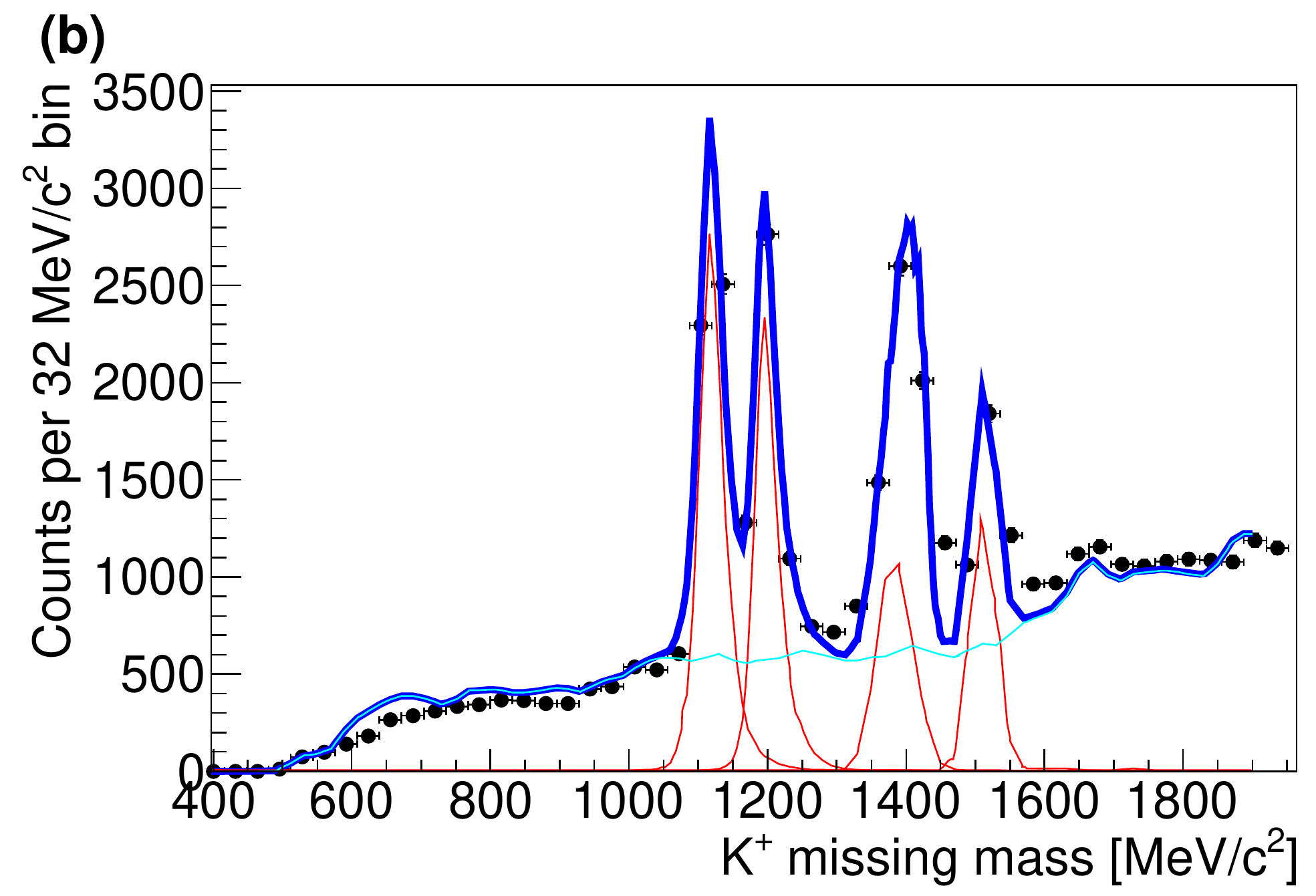}}
	\caption{Missing mass recoiling from forward $K^+$.  The spectrum is fitted with real data describing background from $e^+$ and $\pi^+$ (cyan line) and peaks corresponding to $K^+\Lambda$, $K^+\Sigma^0$, $K^+\Sigma^0$(1385)/$\Lambda$(1405) (nearly mass degenerate) and $K^+\Lambda$(1502) (red lines).}
	\label{fig:forwardkplus}       
\end{figure}

The unique BGO-OD setup allows identification of $K^+Y$ states at extremely forward regions and at very low momentum transfer.  Fig.~\ref{fig:forwardkplus} shows the missing mass from forward $K^+$ for all events where the $K^+$ momentum is below 1~GeV/c.  Peaks corresponding to hyperon masses are immediately visible above a smooth background up to approximately 1900~MeV/c$^2$.  The approximate fit used simulated data to describe the signal, and background arising from misidentified $\pi^+$ and positrons from pair production from the beam.
This kinematic region covers many $\Lambda$ and $\Sigma^*$ states where the evidence of existence is poor or many parameters not accurately determined~\cite{pdg18}.

\begin{figure}[h]
	\centering
	\preliminary{\includegraphics[width=\columnwidth,trim={10cm 0 0.5cm 0.51cm},clip=true]{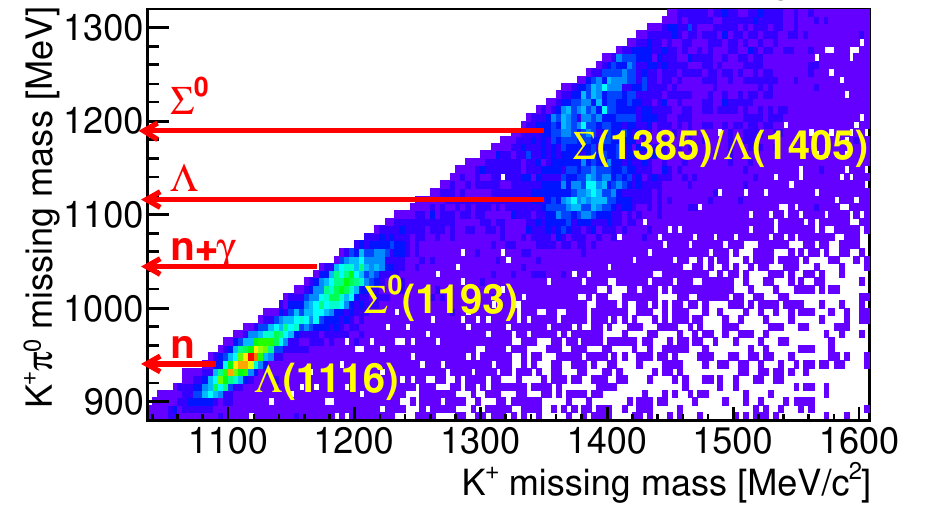}}
	\caption{Missing mass from $K^+\pi^0$ system versus $K^+$ missing mass.
	Yellow labels indicate hyperon mass peaks and red arrows indicated expected missing mass from the $K^+\pi^0$ system.}
	\label{fig:forwardkplus2d}       
\end{figure}

Different hyperon states where the masses overlap can be disentangled via identifying their decays in the BGO Rugby Ball.  Fig.~\ref{fig:forwardkplus2d} shows the same spectra as in Fig.~\ref{fig:forwardkplus} on the x-axis, with an additional axis showing the missing mass from a combined $K^+\pi^0$ system.  This has been used to separate the nearly mass degenerate $\Sigma^0$(1385) and $\Lambda$(1405).

BGO-OD is ideal to investigate the \newline$\gamma p \rightarrow K^+(\Lambda(1405)\rightarrow \Sigma^0\pi^0)$ line shape ($\Lambda(1405)$ invariant mass) and differential cross section.  In addition to providing high statistics data over kinematic regions accessible to other facilities, the kinematics can be extended to very forward, low momentum transfer regions using the forward spectrometer.

 The $\gamma p \rightarrow K^+(\Lambda(1405)\rightarrow \Sigma^0\pi^0)$ signal has been determined by fitting simultaneously to  the two dimensions in Fig.~\ref{fig:forwardkplus2d}, and line shapes and differential cross sections measured~\cite{georgthesis}.  Two distinct analyses were made.  The \textit{Full topology Analysis} has a nearly $4\pi$ angular acceptance, where every particle in the final state $\pi^0\gamma\pi^-p$ was identified and a kinematic fit employed.  The \textit{Forward Spectrometer Analysis} uses the high momentum resolution at forward angles to reconstruct each event from a forward $K^+$ and $\pi^0$ decay only.

\begin{figure}[h]
	\centering
	{\includegraphics[width=\columnwidth,trim={0cm 0 0 0},clip=true]{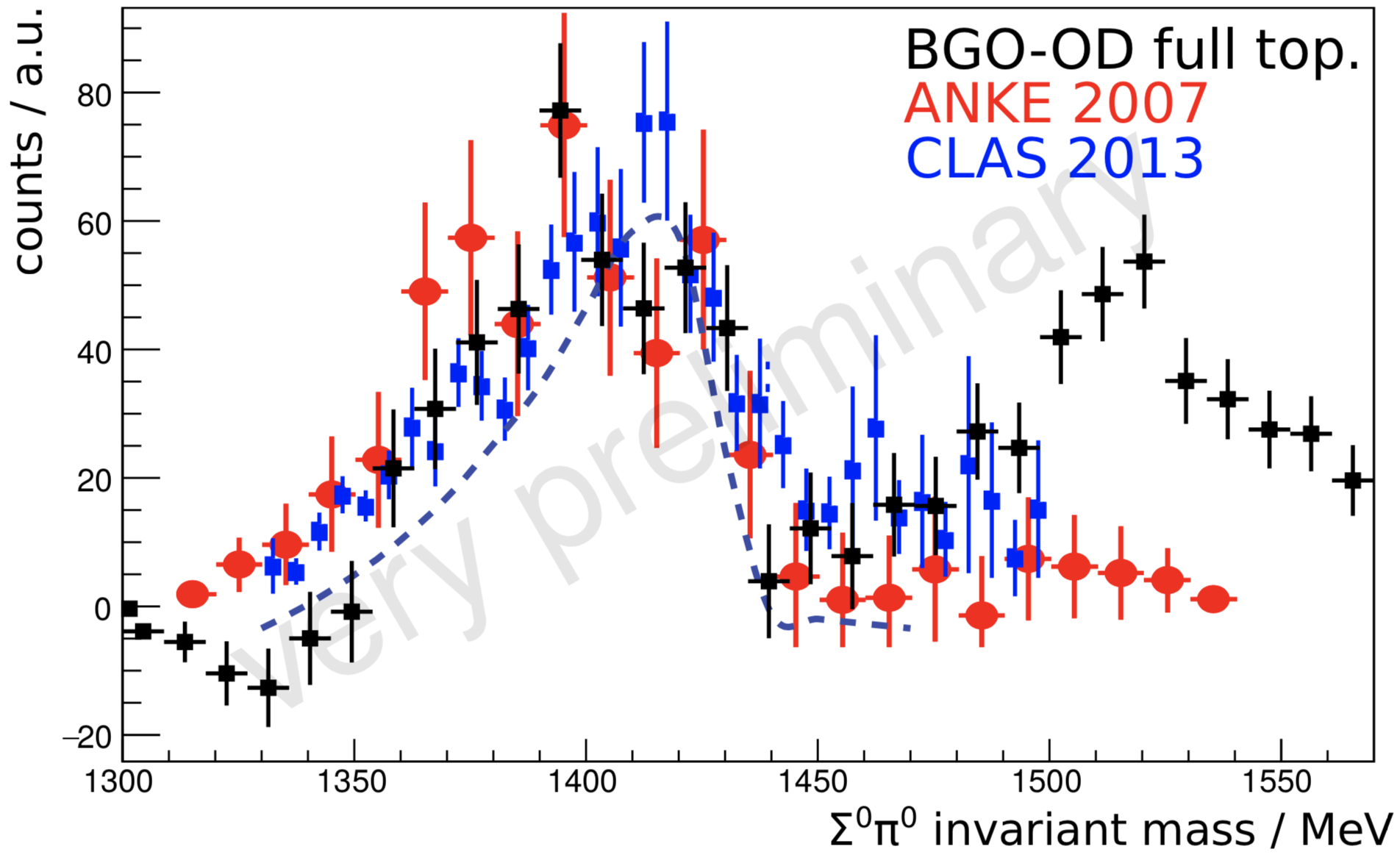}}
	\caption{$\Lambda$(1405)$\rightarrow \pi^0\Sigma^0$ line shape (black points), compared to previous data from Zychor \textit{et al.}~\cite{zychor08} and Moriya \textit{et al.}~\cite{moriya13}  the dashed line is the prediction of Nacher \textit{et al.}~\cite{nacher99}~using a chiral unitary model where the $\Lambda$(1405) is dynamically generated via $K^-p$ interactions.  Analysis by G. Scheluchin~\cite{georgthesis}.}
	\label{fig:l1405lineshape}       
\end{figure}

\begin{figure}[h]
	\centering
	{\includegraphics[width=\columnwidth,trim={0cm 0 0 0},clip=true]{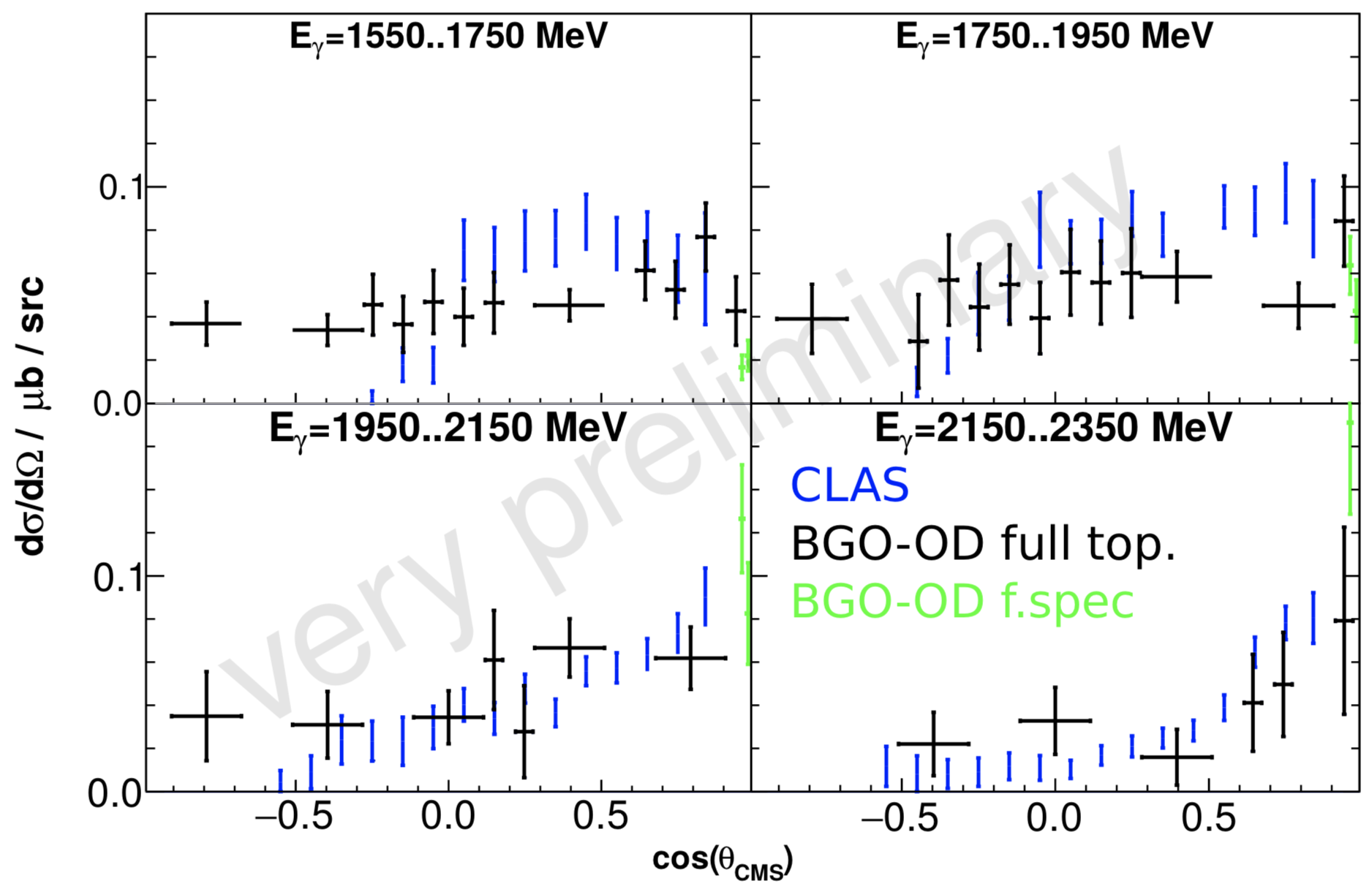}}
	\caption{Preliminary $\gamma p \rightarrow K^+ (\Lambda$(1405)$\rightarrow \pi^0\Sigma^0)$ differential cross section using a full topology and forward $K^+$ analysis (black and green points respectively).  Blue points are previous data of Moriya \textit{et al.}~\cite{moriya13}.    Analysis by G. Scheluchin~\cite{georgthesis}.  }
	\label{fig:l1405cs}       
\end{figure}

\begin{figure}[h]
	\centering
	{\includegraphics[width=\columnwidth,trim={0cm 0 0 0},clip=true]{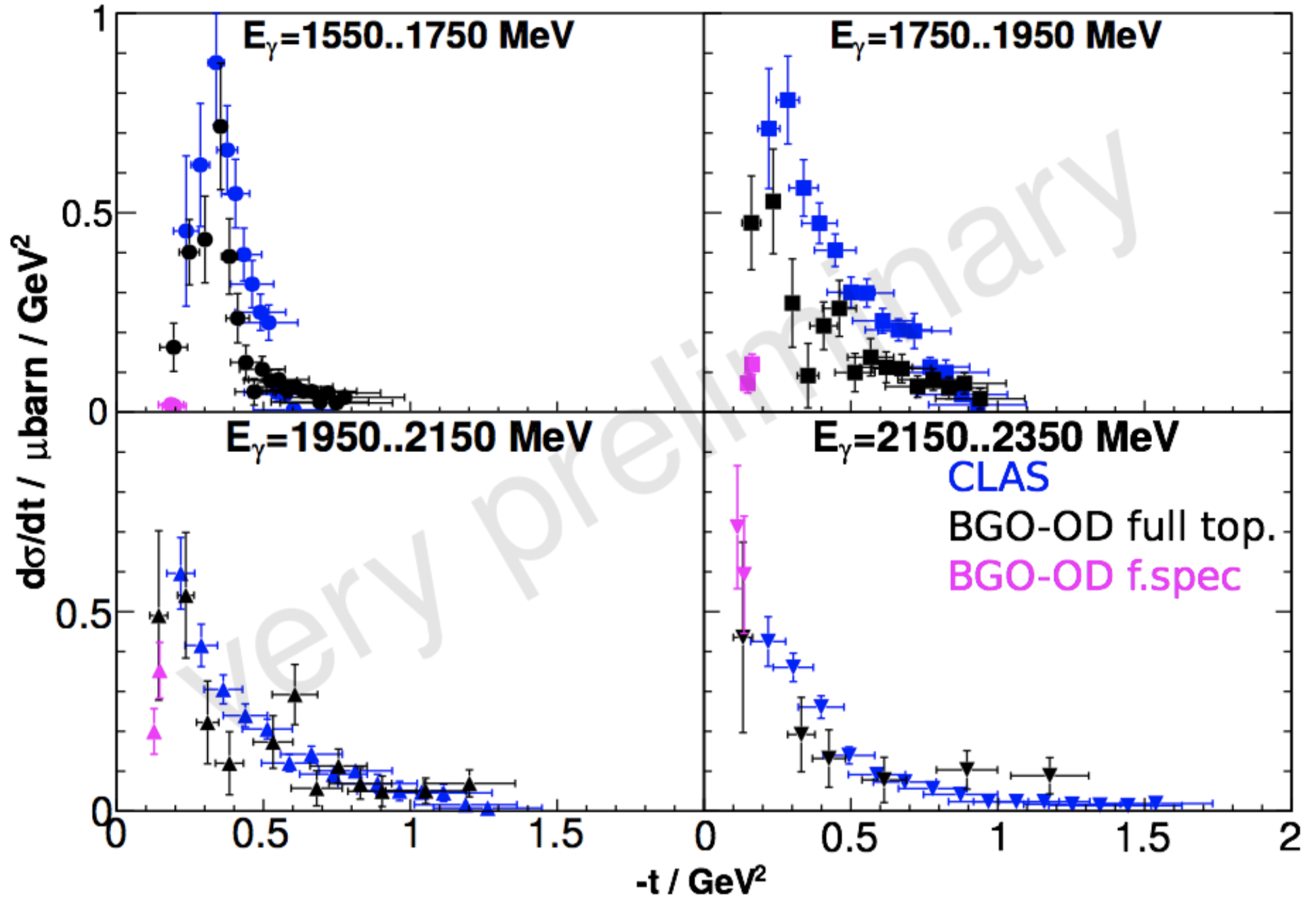}}
	\caption{The same data shown in Fig.~\ref{fig:l1405cs}, but plotted vs $-t$ (black and magenta points show the full topology and forward $K^+$ analysis respectively).  Blue points are previous data of Moriya \textit{et al.}~\cite{moriya13}.   Analysis by G. Scheluchin~\cite{georgthesis}.  }
	\label{fig:l1405csvst}       
\end{figure}

Fig.\ref{fig:l1405lineshape} shows the line shape of the $\Lambda(1405)\rightarrow \pi^0\Sigma^0$ decay for the \textit{Full Topology Analysis}.  The statistics are comparable with the previous data shown and provide a crucial constraint given that BGO-OD is ideally suited to identify the neutral $\Lambda$(1405) decay.  Line shapes at very forward angles using the \textit{Forward Spectrometer Analysis} also show consistent results.  The differential cross section for both analyses is shown in Fig.~\ref{fig:l1405cs}, where extremely forward angles can be accessed for the first time.  This is also demonstrated in Fig.~\ref{fig:l1405csvst}, where the data are converted into a differential cross section with respect to the square of the momentum exchange, $-t$.  See Ref.~\cite{georgNSTAR} for further details.

\subsection{Forward $K^+\Lambda$ and $K^+\Sigma^0$ photoproduction}

Ground state, $K^+\Lambda$ and $K^+\Sigma^0$ photoproduction is virtually unconstrained at forward angles.
The lack of data and discrepancies in the existing world data prevents accurate descriptions by isobar models and partial wave analyses, where many solutions have large deviations (see for example, Ref.~\cite{skoupil18}).  Data at this very low momentum exchange region is also crucial in constraining the elementary reaction processes in hypernuclei electroproduction, where the required low $Q^2$ is comparable to the photoproduction process.
The excellent forward acceptance and high momentum resolution allows BGO-OD to resolve these issues.

\begin{figure}[h]
	\centering
	\preliminary{\includegraphics[width=\columnwidth,trim={0cm 0 0 0},clip=true]{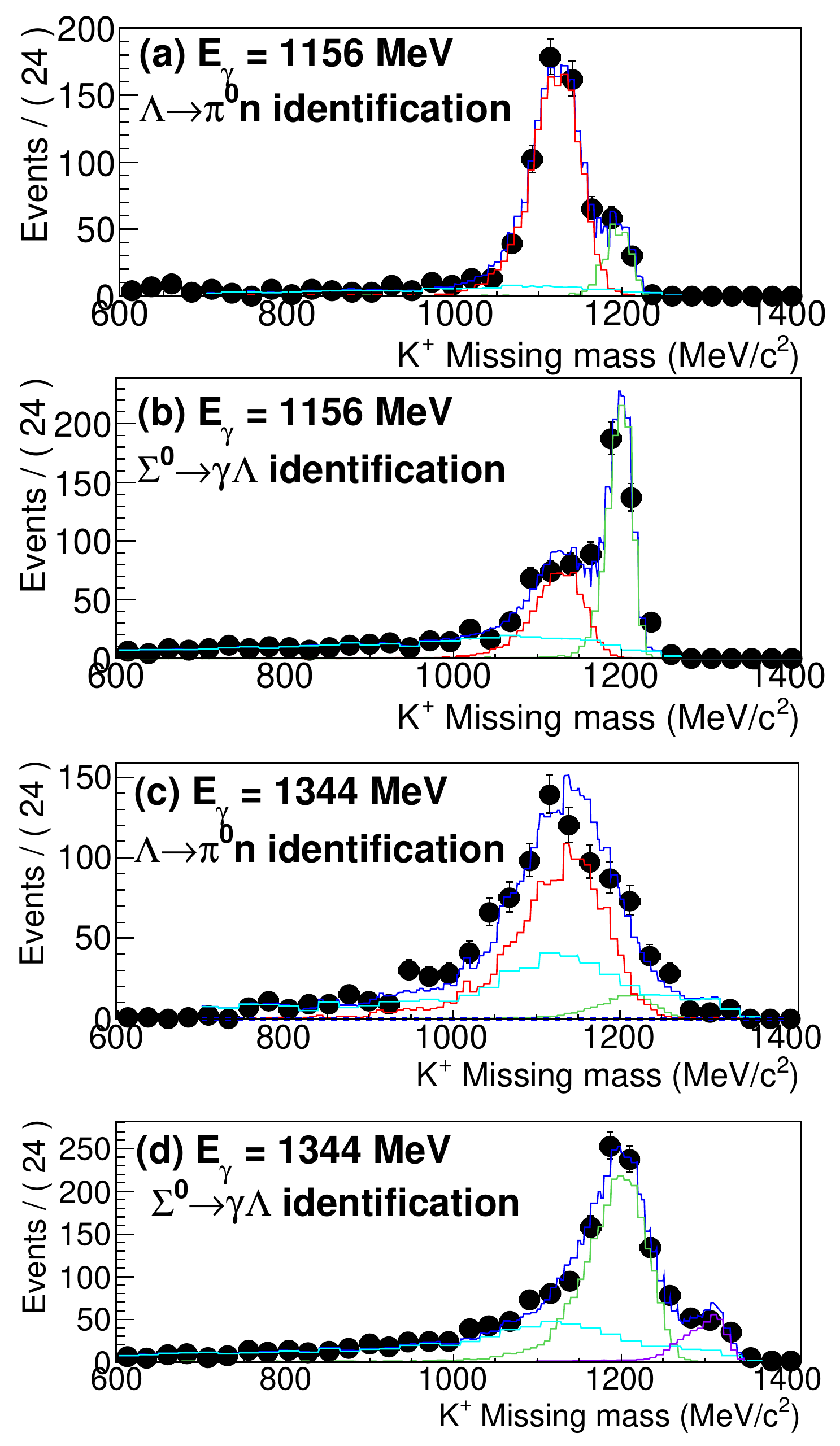}}
	\caption{Missing mass recoiling from forward $K^+$ candidates (black points) for two different photon beam energies and selection criteria to either accentuate the $K^+\Lambda$ or $K^+\Sigma^0$ signal.  The spectra are fitted with $e^+$ background (described in the text), simulated $K^+\Lambda$, $K^+\Sigma^0$ and $K^+\Sigma^0(1385)$ (cyan, red, green and purple lines respectively). The blue line is the summed fit.  }
	\label{fig:kaonmm}       
\end{figure}

Additional selection criteria were used to enhance the signal above background for $K^+\Lambda$ and $K^+\Sigma^0$ photoproduction.  
For $K^+\Lambda$ photoproduction, a diagonal selection cut was applied to the two dimensional plot in Fig.~\ref{fig:forwardkplus2d} to select the $\Lambda\rightarrow \pi^0 n$ decay.
For $K^+\Sigma^0$ photoproduction, the $\Sigma^0$ decay photon was identified in the same manner as described in section~\ref{sec:k0}.  
Fig.~\ref{fig:kaonmm} shows the missing mass for forward $K^+$ (corresponding to $\cos\theta_{CM}^{K^+} > 0.9$) for two different photon beam intervals.  The two selection criteria enhance the required channel.  The non-$K^+$ background indicated by the cyan line in the figure mostly originates from positrons from pair production in the beam.  This distribution was determined using an equivalent analysis for negatively charged particles, where the electron distribution was identical.  The largest contribution of hadronic reaction background was determined as $\gamma p \rightarrow \pi^0 \Delta^+$, where a $\pi^+$ from the $\Delta^+$ decay was misidentified as a $K^+$.  Simulated data were used to determine systematic effects of this upon the extracted signal yields.

\begin{figure}[h]
	\centering
	\preliminary{\includegraphics[width=\columnwidth,trim={0cm 0 2cm 0},clip=true]{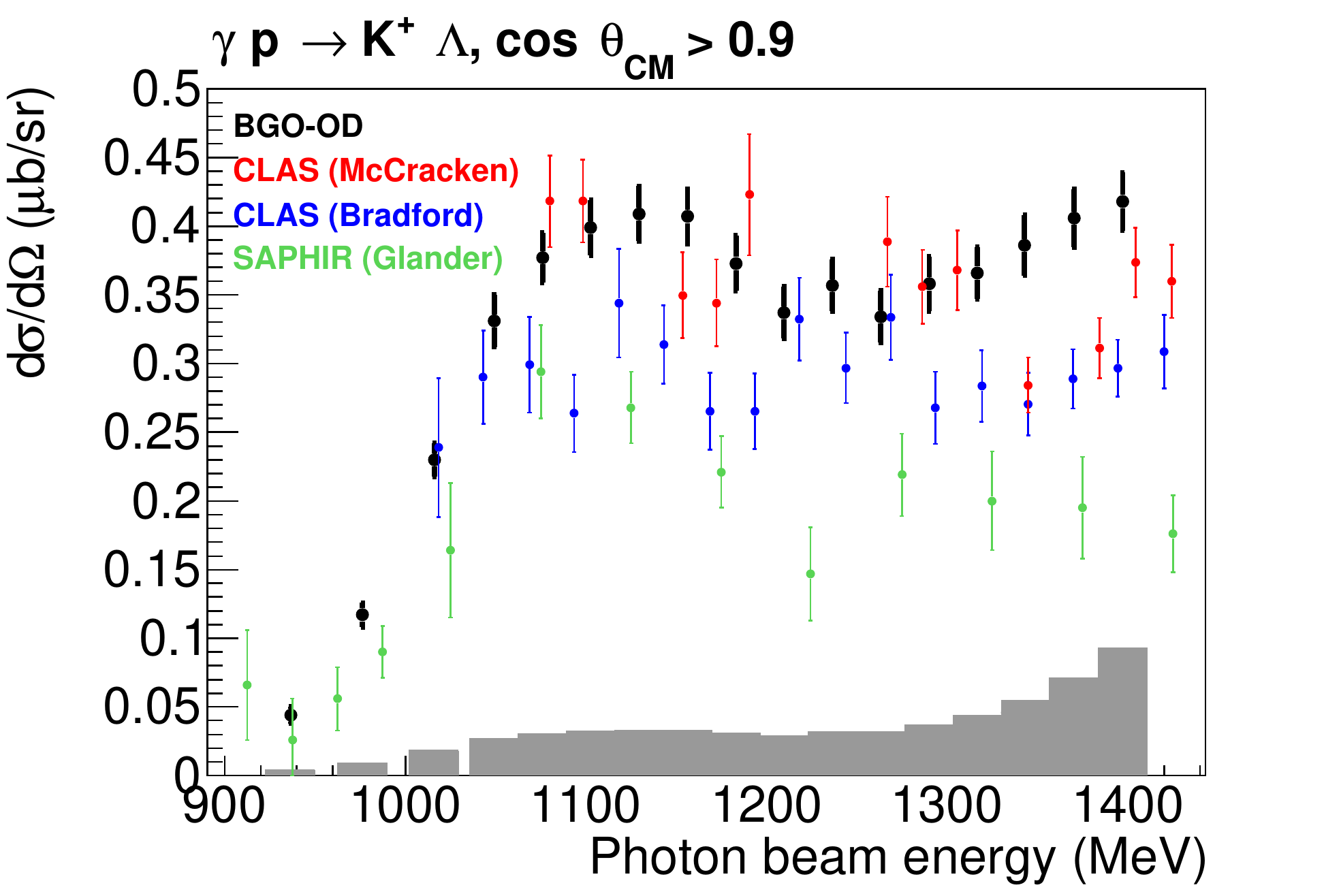}}
	\caption{$\gamma p \rightarrow K^{+}\Lambda$ differential cross section for $\cos\theta_{CM} > 0.9$ (Black points).  Systematic errors are the filled grey bars on the abscissa.  Previous data of McCracken~\cite{mccracken10}, Bradford~\cite{bradford06} and Glander~\cite{glander04} shown in red, blue and green respectively.  The data of McCracken and Bradford are at the more backward interval, $0.85 < \cos\theta_{CM} < 0.95$.}
	\label{fig:klcs}       
\end{figure}

Fig.~\ref{fig:klcs} shows the $\gamma p \rightarrow K^+\Lambda$ differential cross section for $\cos\theta_{CM}^{K^+} > 0.9$.  The statistical error is significantly smaller than previous datasets, allowing for the first time an accurate description at small angles, low momentum transfer regions where $t$-channel mechanisms dominate the reaction process.  The high statistics allow a discrimination between previous datasets which exhibit discrepancies.  It should be noted that the CLAS data are at the slightly more backward angle interval of \newline $0.85 < \cos\theta_{CM}^{K^+} < 0.95$. The dominant systematic errors shown on the abscissa are the flux normalisation and beam spot position (4\% and 5\% respectively) up to approximately 1260~MeV photon beam energies.  At higher energies the fitting to the signal and background is the dominant error (approximately 12\%).
	
\begin{figure}[h]
	\centering
	\preliminary{\includegraphics[width=\columnwidth,trim={0cm 0 2cm 1cm},clip=true]{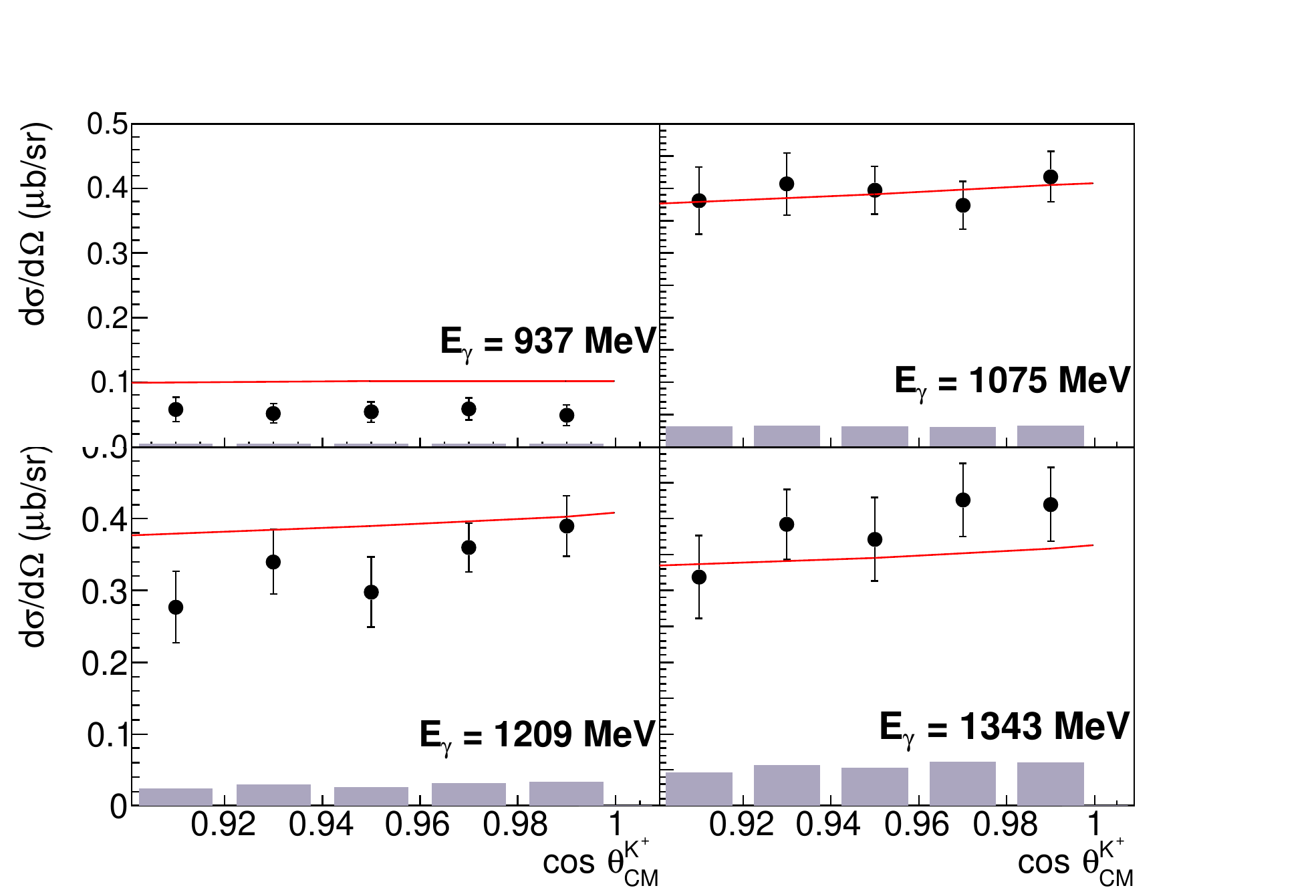}}
	\caption{$\gamma p \rightarrow K^{+}\Lambda$ differential cross section for four example photon beam energies (labelled inset) versus $\cos_{CM}^{K^+}$.  Systematic errors are the filled grey bars on the abscissa.   The red line is the Bonn-Gatchina PWA solution~\cite{anisovich14}.}
	\label{fig:klcsangle}       
\end{figure}

BGO-OD is able to measure this forward region in fine polar angle intervals.  Fig.~\ref{fig:klcsangle} shows four examples of the angular distribution of the differential cross section, enabling accurate extrapolations to the minimum possible momentum exchange at the most forward angle.

The recoil polarisation, $P_\Lambda$ was accessed via the self-analysing weak decay of the $\Lambda$, given in Eq.~\ref{eq:recpol}, where $N_{\uparrow/\downarrow}$ are the $\pi^0$ yield above or below the reaction plane in the $\Lambda$ rest frame and the weak decay parameter, $\alpha = 0.642$~\cite{pdg18}.

\begin{equation}\label{eq:recpol}
P_\Lambda = \frac{2}{\alpha}\frac{N_\uparrow - N_\downarrow}{N_\uparrow + N_\downarrow}
\end{equation}

\begin{figure}[h]
	\centering
	\preliminary{\includegraphics[width=\columnwidth,trim={0cm 0 0cm 0cm},clip=true]{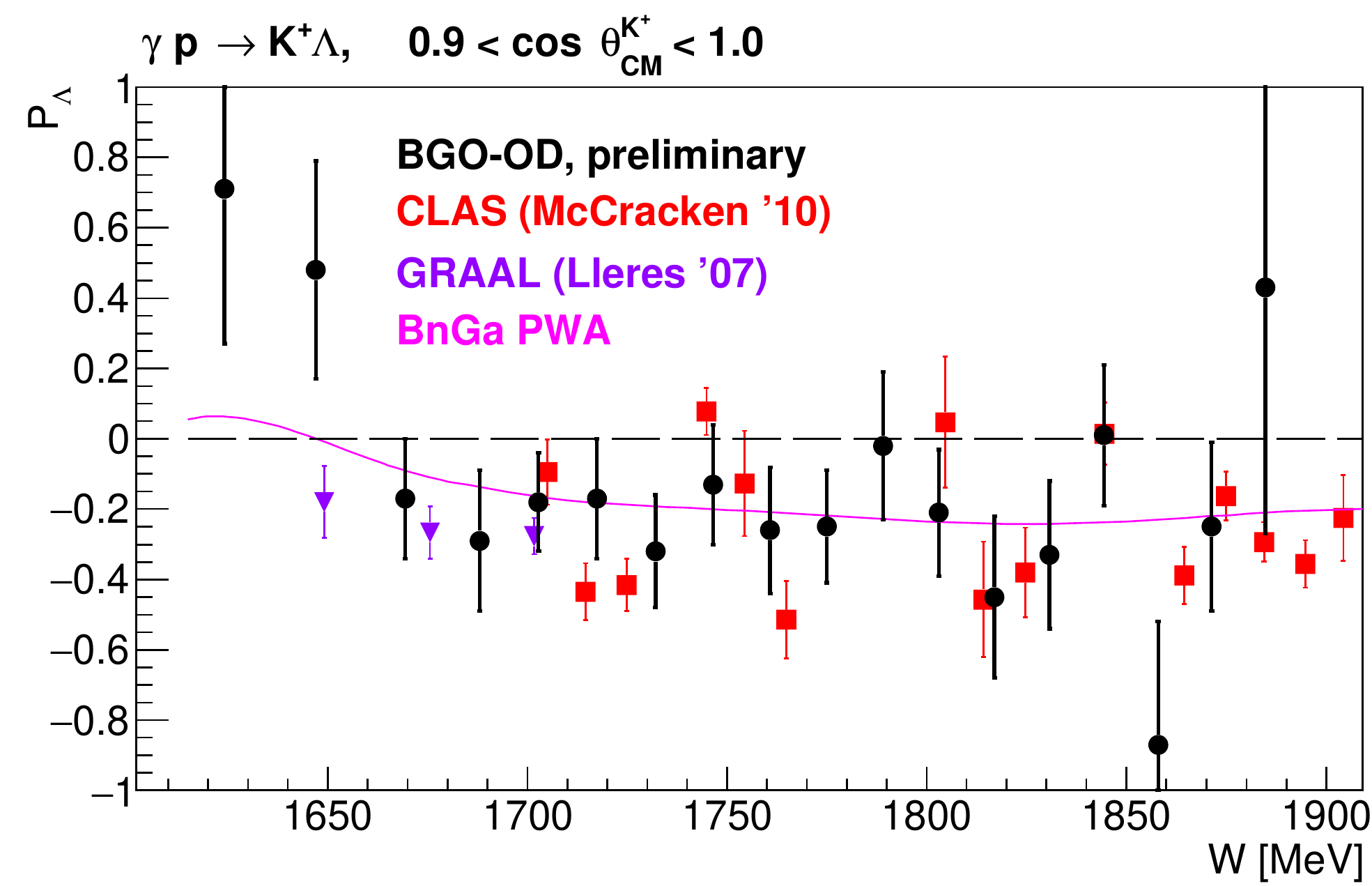}}
	\caption{$\gamma p \rightarrow K^{+}\Lambda$ recoil polarisation for $\cos_{CM}^{K^+} > 0.9$ (black points).  Previous data shown from McCracken \textit{et al.}~\cite{mccracken10}, Lleres \textit{et al.}~\cite{lleres07} and the Bonn-Gatchina PWA solution~\cite{anisovich07} shown in red triangles, purple triangles and magenta line respectively.}
	\label{fig:recpol}       
\end{figure}

Fig.~\ref{fig:recpol} shows the recoil polarisation for $\cos\theta_{CM}^{K^+} > 0.9$.  The BGO-OD data are consistent with the previous data shown.

\begin{figure}[h]
	\centering
	\preliminary{\includegraphics[width=\columnwidth,trim={0cm 0 2cm 0cm},clip=true]{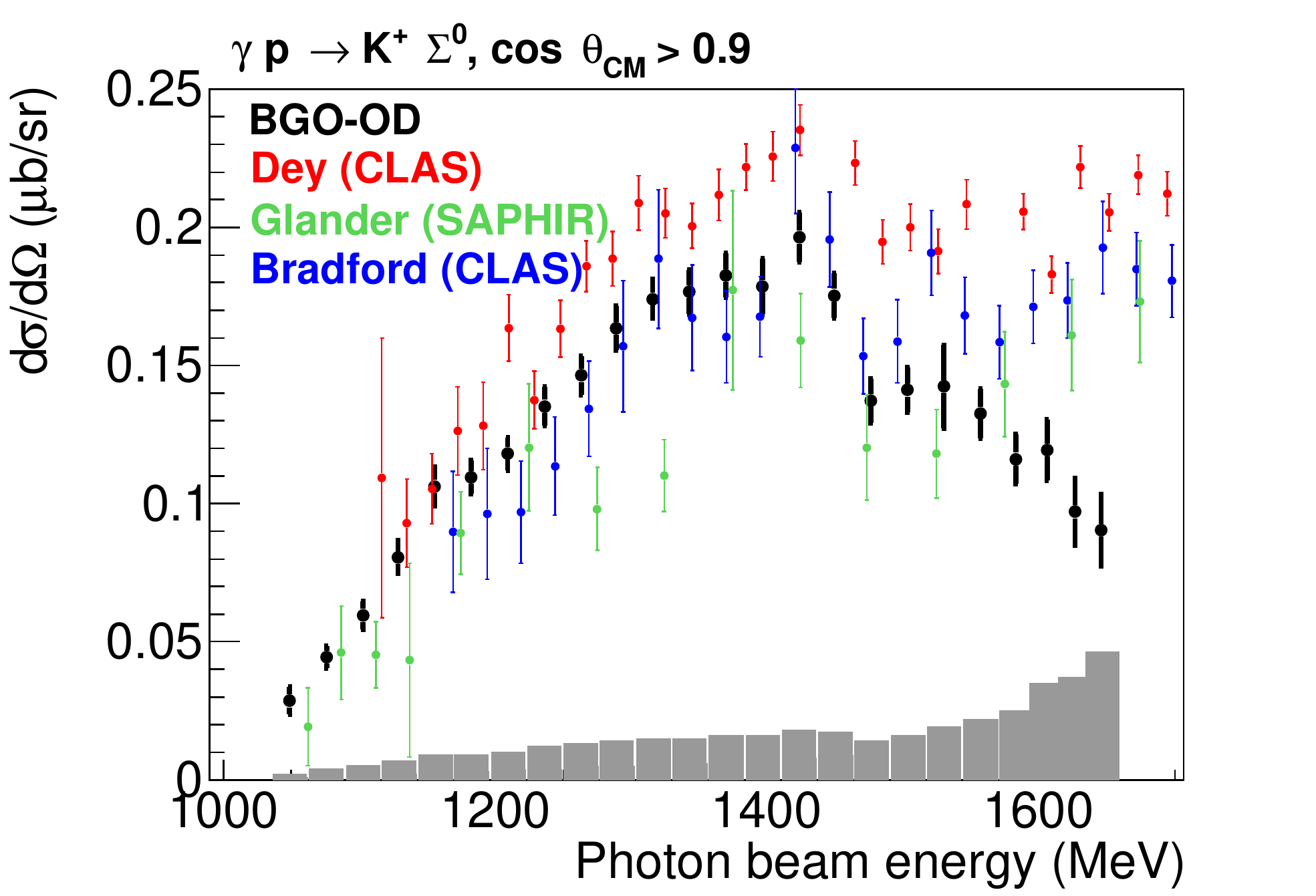}}
	\caption{$\gamma p \rightarrow K^{+}\Sigma^0$ differential cross section for  \newline $\cos\theta_{CM} > 0.9$ (Black points).  Systematic errors are the filled grey bars on the abscissa.  Previous data of Dey~\cite{dey10}, Glander~\cite{glander04} and Bradford~\cite{bradford06}  shown in red, blue and green respectively.  The data of Dey and Bradford are at the more backward interval, $0.85 < \cos\theta_{CM} < 0.95$.}
	\label{fig:kscs}       
\end{figure}

The differential cross section for $\gamma p \rightarrow K^+\Sigma^0$ for $\cos\theta_{CM} > 0.9$ is shown in Fig.~\ref{fig:kscs}.  As for $K^+\Lambda$, the statistics are of unprecedented accuracy at this forward angle.
The data suggest a "cusp-like" structure at a photon beam energy of approximately 1470~MeV where there is a drop in the cross section of approximately 25\%.  Previous data sets supported this however with insufficient statistics for a clear statement.

\begin{figure}[h]
	\centering
	\preliminary{\includegraphics[width=\columnwidth,trim={0cm 0 0cm 0},clip=true]{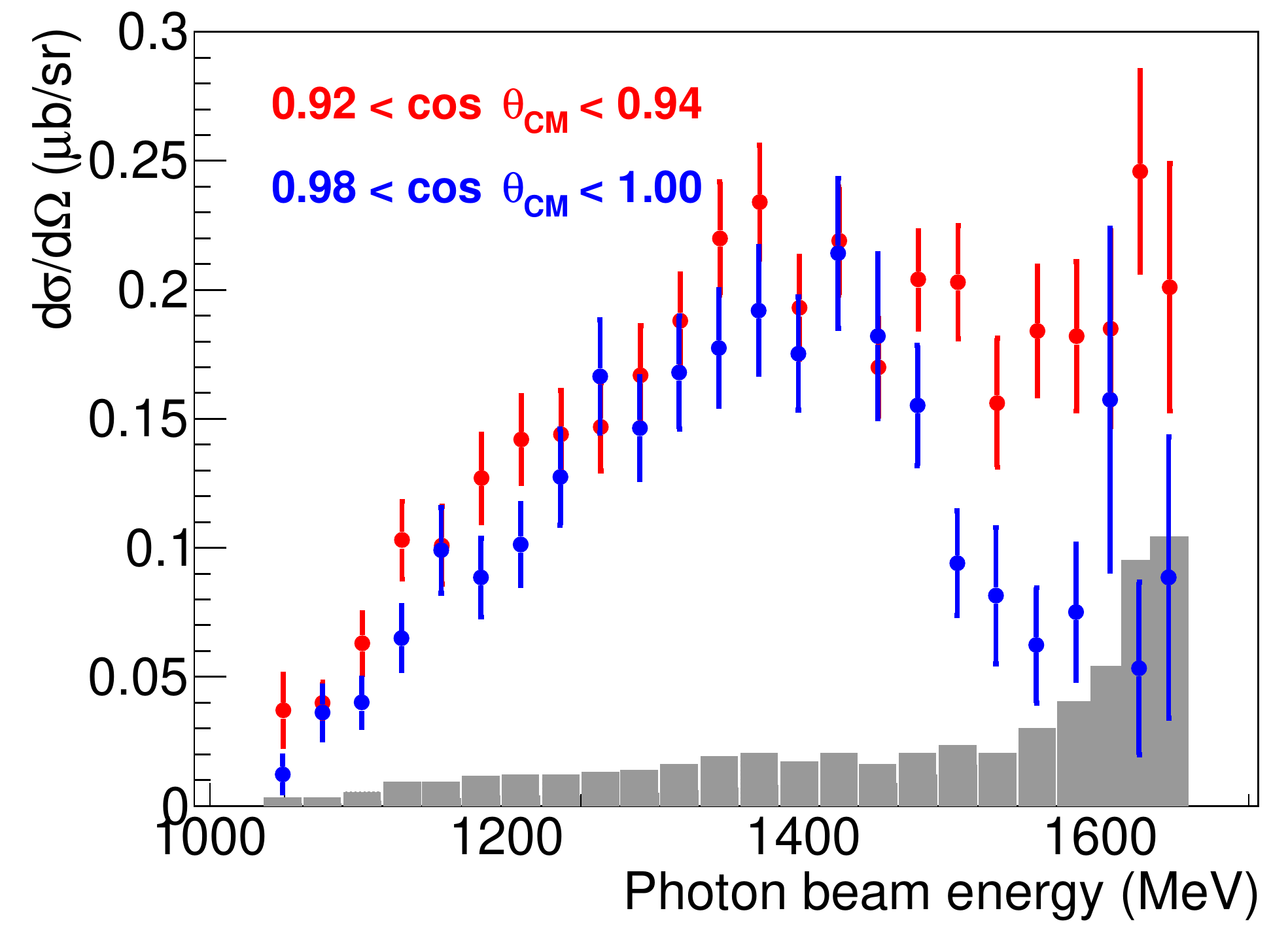}}
	\caption{$\gamma p \rightarrow K^{+}\Sigma^0$ differential cross section for \newline $0.92 < \cos\theta_{CM} < 0.94$ and  $\cos\theta_{CM} > 0.98$ (red and blue points respectively).  Systematic errors for the more backward angle interval are the filled grey bars on the abscissa.}
	\label{fig:kscscusp}       
\end{figure}

Similarly to $K^+\Lambda$ photoproduction, the data were divided into fine, 0.02 $\cos_{CM}^{K^+}$ intervals, two examples of which are shown in Fig.~\ref{fig:kscscusp}.  The "cusp-like" structure appears more pronounced at the most forward, lowest momentum exchange region.  This appears consistent with the expected behaviour of a molecular like system or sub-threshold effects playing important roles at very low momentum exchange regions.  This is close to multiple thresholds of open and hidden strangeness, notably $K^+\Lambda$(1405), $f_0 p$ (where the $f_0$ has been previously described as a $K\bar{K}$ molecule) and $\eta' p$, however mechanisms responsible are still under discussion.

\section{Conclusions}


The BGO-OD experiment is an ideal setup to investigate associated strangeness photoproduction at very forward angles and low momentum transfer where exotic behaviour may manifest.  The first key results vindicate this argument, with a cusp in the $K^+\Sigma^0$ photoproduction cross section observed at very forward angles, and a peak in $K^0\Sigma^0$  photoproduction cross section at the $K^*$ thresholds.  The $\Lambda(1405)\rightarrow \pi^0\Sigma^0$ line shape and differential cross section is also measured over a broad kinematic region, with high statistics and extending to very low momentum transfer regions.

\subsection*{Acknowledgements}
It is a pleasure to thank the ELSA staff for a reliable and
stable operation of the accelerator and the technical staff
of the contributing institutions for essential help in the
realisation and maintenance of the apparatus.

This work was supported by the Deutsche Forschungsgemeinschaft project numbers 388979758 and 405882627.  Our Russian collaborators thank the Russian Scientific Foundation (grant RSF number 19-42-04132) for financial support.

%
%
%


\end{document}